# Molecular Beam Epitaxy Growth of High Crystalline Quality LiNbO$_3$

M. Brooks Tellekamp[a], Joshua C. Shank[a], Mark S. Goorsky[b], and W. Alan Doolittle[a,1]

[a] Department of Electrical and Computer Engineering, Georgia Institute of Technology, 777 Atlantic Drive, Atlanta Ga 30332, USA
[b] Department of Materials Science and Engineering, University of California, 410 Westwood Plaza, 3111 Engineering V, Los Angeles, Ca 90095, USA

**Abstract—** **Lithium niobate is a multi-functional material with wide reaching applications in acoustics, optics, and electronics. Commercial applications for lithium niobate require high crystalline quality currently limited to bulk and ion sliced material. Thin film lithium niobate is an attractive option for a variety of integrated devices, but the research effort has been stagnant due to poor material quality. Both lattice matched and mismatched lithium niobate are grown by molecular beam epitaxy (MBE) and studied to understand the role of substrate and temperature on nucleation conditions and material quality. Growth on sapphire produces partially coalesced columnar grains with atomically flat plateaus and no twin planes. A symmetric rocking curve shows a narrow linewidth with a full width at half-maximum (FWHM) of 8.6 arcsec (0.0024°) which is comparable to the 5.8 arcsec rocking curve FWHM of the substrate, while the film asymmetric rocking curve is 510 arcsec FWHM. These values indicate that the individual grains are relatively free of long-range disorder detectable by x-ray diffraction (XRD) with minimal measurable tilt and twist and represents the highest structural quality epitaxial material grown on lattice mismatched sapphire without twin planes. Lithium niobate is also grown on lithium tantalate producing high quality coalesced material without twin planes and with a symmetric rocking curve of 193 arcsec, which is nearly equal to the substrate rocking curve of 194 arcsec. The surface morphology of lithium niobate on lithium tantalate is shown to be atomically flat by atomic force microscopy (AFM).**

## I. INTRODUCTION

Lithium niobate (LiNbO$_3$) is a well-studied material with a wide range of applications in acoustics, optics, and electronics due to a long list of multi-functional properties. Lithium niobate is piezoelectric, pyroelectric, ferroelectric, photoelastic, birefringent, photorefractive, and exhibits the linear electro-optic effect (Pockels effect). Of these properties, the piezoelectric, pyroelectric, electro-optic, and photoelastic effects have large associated coefficients and therefore exaggerated effects [1,2]. These favorable properties have made lithium niobate an important material for waveguides and acousto-optic devices, where it has found use as surface acoustic wave (SAW) devices [3], acousto-optic modulators [4], electro-optic modulators [5], waveguides [6], and second-harmonic generators [7] among various other devices. Producing this material in thin film form gives multiple advantages compared to bulk material including monolithic structures and heterojunctions for complex integrated devices, higher index mismatch waveguides for tighter wave confinement and lower evanescent losses, lower required drive voltages, and the possibility of enhanced and engineered properties by doping and strain.

[1] alan.doolittle@ece.gatech.edu; *Tele.&Fax:* 404.894.9884

Lithium niobate is a mature technology when considering bulk Czochralski grown material, but epitaxial thin films have yet to make commercial impact as research was largely abandoned in the late 1990s and early 2000s in favor of ion sliced or "smart cut" lithium niobate [8]. Ion slicing is a method of cleaving a bulk wafer along a plane weakened by $H^+$ or $He^+$ ion implantation, completed by wafer bonding and transfer. While ion sliced lithium niobate has been successful in providing "thin film" materials, it has significant limitations. Ion sliced material is limited to thicknesses greater than ≈ 500 nm, it must be polished to improve the surface roughness, it cannot be centrally incorporated into heterostructures, and it must be annealed at high temperatures to regain the electro- and nonlinear optical properties [8].

Epitaxial lithium niobate will be commercially viable only if the crystal quality can be improved to minimize optical and acoustic losses to acceptable values, especially if it can be grown on a low index material such as sapphire. Previous techniques used to grow thin film lithium niobate include sputtering [9–13], sol-gel methods [14], pulsed laser deposition (PLD) [15–18], liquid phase epitaxy (LPE) [19,20], chemical vapor deposition (CVD) [21–24], and molecular beam epitaxy (MBE) [25–29] on substrates including $LiNbO_3$, $LiTaO_3$, $Al_2O_3$, and SiC. Most of these methods produce material with 60° rotational domains and large symmetric rocking curves. Rotational domains are undesirable in optical devices due to scattering at domain boundaries. LPE produces material of nearly bulk quality with symmetric rocking curves of 11 arcsec, but in-plane structural characterization was not reported [19,20]. PLD grown material has also shown near-bulk rocking curves of 13 arcsec but reported rotational domains [18]. Other PLD grown material is single domain, but with a larger symmetric rocking curve of 612 arcsec [15]. Sputtered material shows higher overall rocking curve values and rotational domains; only one example shows minimal but still present twin planes [30]. Feigelson and Lee have produced the best overall material by CVD with rocking curves of 158 arcsec on sapphire substrates and 36 arcsec on lithium tantalate, where the sapphire based films show minimal twin planes which are removed by a 750°C 24 h anneal and the lithium tantalate based films do not show twin planes [21,22].

Previous MBE grown material is poorly reported in general with only one reference analyzing rotational domains [29] which were present, and two quoting symmetric rocking curves of 936 and 150 arcsec, both grown on SiC [26,28]. There is no discussion of surface morphology for MBE grown material in current literature. In order to develop thin film heterostructures and devices which take advantage of the multi-functional properties of lithium niobate it is desirable to understand the film nucleation and coalescence of MBE grown material and improve crystal quality.

## II. EXPERIMENT

### A. Growth System

The films grown in this work are produced in a Varian Gen-II MBE system which has been customized for the growth of lithium niobium oxides using a lithium assisted metal-halide growth chemistry discussed in detail by *Henderson et al.* [31]. The method involves the sublimation of $NbCl_5$, present as the dimer $Nb_2Cl_{10}$, from a custom water and filament heated near-ambient cell [32]. The custom cell consists of separate bulk and tip zones along with an aperture plate that limits conductance. In this way, stable temperatures and flux can be achieved over the course of a growth where substrate temperature radiation can significantly increase the vapor flux of the low temperature cell (typical bulk operating temperatures are 35–50 °C).

A modified Veeco corrosive series antimony cracker with a custom solid tantalum crucible is used to supply lithium to the system [31]. Lithium is used to getter the chlorine from the dimer in the presence of a hot substrate, desorbing from the surface as LiCl and leaving behind atomic niobium. Excess lithium may be supplied to incorporate into the film, and in the presence of molecular oxygen gas can form a variety of lithium-niobium-oxygen phases such as $LiNbO_2$, $LiNbO_3$, $Li_3NbO_4$, and $LiNb_3O_8$. The phase which is grown depends on the Li:Nb flux ratio, the supplied oxygen, and the substrate temperature. The $LiNbO_3$ films in this work are grown with a $Li:NbCl_5$ flux ratio ≈ 0.7 as measured by a Bayard-Alpert style ionization gauge and at a substrate temperature of 1000–1025°C. Substrate temperatures less than 1000°C do not produce single phase $LiNbO_3$ films.

A custom substrate heater is used in this MBE system to allow growth at high temperatures in a corrosive oxygen and chlorine environment. The heater consists of a thick tantalum filament encased in pyrolytic boron nitride. The substrate can be heated to 1000°C in an oxygen background of $6.67 \times 10^{-6}$ kPa.

### B. Substrates

Lithium niobate is a trigonal crystal in the r3c space group with a unit cell comprised of oxygen octahedra with interstices that are one third lithium, one third niobium, and one third vacant [1]. The hexagonal oxygen sublattice forms layers in the basal plane, and in the ferroelectric phase the lithium atoms are positioned 0.37 Å above or below these layers, with both sites equally energetically favorable [1]. This offset produces the polar and ferroelectric nature of lithium niobate. A sufficient applied field shifts the lithium atoms across the oxygen layer switching the polarity [1].

The hexagonal relationship in the crystal structure that will be exploited for epitaxy on sapphire is not the supercell based hexagon a = 5.148 Å but the pseudo-hexagonal average oxygen sublattice a = 2.986 Å which is 30 degrees rotated from the supercell. The three distinct atomic distances are $a_1$ = 2.718 Å, $a_2$ = 2.878 Å, and $a_3$ = 3.363 Å. This distinction is illustrated in Fig 1.

In order to understand the role of the substrate on film quality and nucleation, films in this work were grown on both lattice matched and mismatched substrates. The lattice parameters and coefficient of thermal expansion (CTE) of selected substrates and nucleation layers are given in Table I. For lattice mismatched growth, c-plane (006) sapphire was used, and for lattice matched growth c-plane $LiTaO_3$ was used which has been shown to be more vacuum stable than $LiNbO_3$ [33,34].

The 1 cm square substrates are backside metallized with ~2 μm chrome followed by ~2 μm chrome oxide in a Denton Discovery reactive sputtering system. The substrates are then cleaned by solvents followed by $H_2SO_4$:$H_2O_2$ in a 4:1 ratio at 120°C before loading into an introductory vacuum chamber and degassed at 200°C for 1 h.

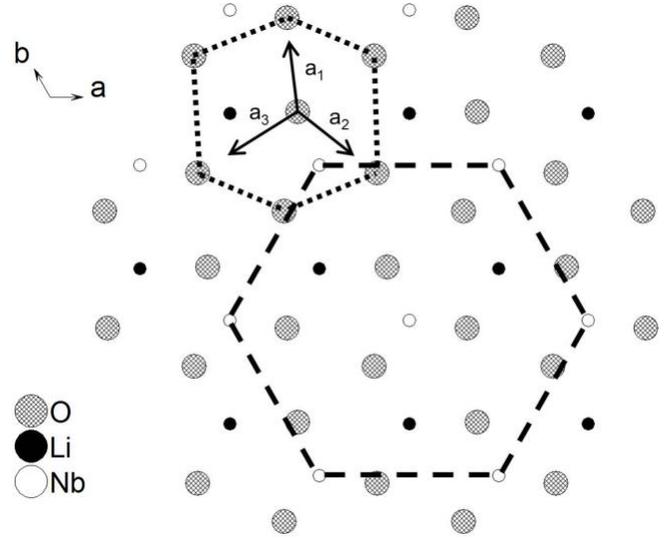

**Fig 1.** The c- plane of lithium niobate (c-axis out of plane) showing the hexagonal supercell as dashes (a = 5.148 Å) and the 30° rotated pseudo-hexagon in the oxygen sublattice as dots (a = 2.986 Å). The three weighted pseudo-hexagonal parameters are notated $a_1$, $a_2$, and $a_3$.

**Table I.** Selected substrates and their properties for $LiNbO_3$ epitaxy.

| Material | pseudo-hexagonal a-spacing (Å) | CTE (× $10^{-6}$ /°C) |
| --- | --- | --- |
| $LiNbO_3$ | 2.98 | 16.2 [35] |
| $LiNbO_2$ | 2.92 (2.2%) | Unknown |
| $LiTaO_3$ | 2.99 (0.1%) | 15.4 [35] |
| $Al_2O_3$ | 2.74 (8.7%) | 5.2 [36] |
| SiC (6H) | 3.08 (3%) | 3.27 [37] |
| Nb (BCC) | 2.86 (4.4%) | 7.3 [38] |

### C. Characterization

The films in this work were structurally characterized by x-ray diffraction (XRD) using a Phillips X'pert Pro MRD with a hybrid parabolic mirror/Ge 4-bounce crystal monochromator delivering Cu-K$\alpha_1$ radiation (λ = 1.54056 Å) in both double- and triple-axis geometries. For double axis measurements, a 0.0625° slit was used on the incident optics and a 0.25° slit was used on the detector optics. For triple-axis measurements no slits were used on the incident optics. Symmetric scans were used to determine c-spacing strain and uniformity, single phase nature, tilt angle, and thickness when applicable. Asymmetric scans were used to analyze films for rotational domains and twist.

The surface morphology of the lithium niobate films was analyzed by a Veeco atomic force microscopy (AFM) system in tapping mode using silicon probes with a tip radius of 7 nm and theoretical horizontal resolution of 5 nm.

## III. RESULTS

### A. Lattice Mismatched Substrate

$LiNbO_3$ thin films were grown on c-plane $Al_2O_3$ substrates to analyze the effects of lattice mismatch on nucleation, structure, and morphology. $Al_2O_3$ has a pseudo-hexagonal structure with a weighted lattice constant shown in Table I giving a mismatch of 8.7% with $LiNbO_3$. Additionally, the CTE of $LiNbO_3$ is nearly 3 times larger than that of sapphire. Because of this large lattice mismatch, films are expected to relax quickly during growth and undergo tensile strain during cool down due to the CTE mismatch.

Fig 2 shows an x-ray diffractogram of single phase c-oriented $LiNbO_3$ grown on c-oriented sapphire at a substrate temperature of 1000°C. The film shows Pendellösung fringes which give a thickness of 57 nm and indicate a smooth surface and interface. It was observed that the rocking curve of the (006) reflection could only be resolved in the triple-axis configuration. Fig 3a shows the triple-axis measured symmetric rocking curve with a wider low intensity portion and a higher intensity peak in the center. The low intensity peak was fit to a FWHM of 121 arcsec, and the higher intensity central peak has a narrow FWHM of 8.6 arcsec (0.0024°). This value approaches the theoretical minimum resolvable peak for the XRD system in the triple-axis configuration; however, the peak shows ~30 data points, which can be seen in the inset of Fig 3a, showing that the value is well resolved. Fig 3b shows the symmetric (006) rocking curve of the $Al_2O_3$ substrate which contains the same features observed in the film rocking curve. The narrow peak is 5.8 arcsec FWHM with a wider low intensity portion fit to 67 arcsec FWHM. The coincidence of these features implies the $LiNbO_3$ tilt uniformity is matched to that of the $Al_2O_3$ substrate with slightly titled grains appearing in both substrate and film.

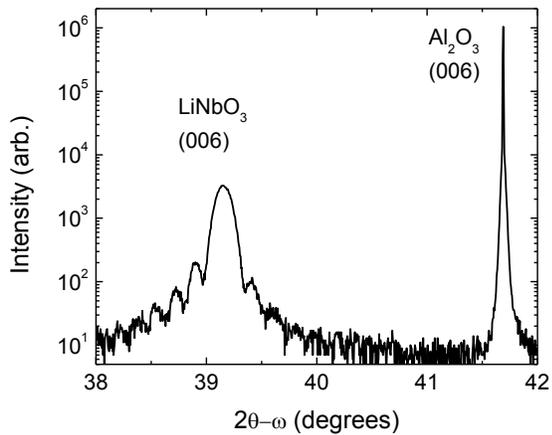

**Fig 2.** XRD 2θ-ω scan of c-oriented LiNbO$_3$ grown on c-oriented sapphire (*) at a peak 2θ-ω value 0.43% strained tensile to the bulk value. Pendellösung fringes indicate a thickness of 57 nm.

An asymmetric rocking curve of the LiNbO$_3$ (012) reflection was also obtained in a skew geometry. The FWHM of this asymmetric peak is 510 arcsec. The asymmetric rocking curve FWHM is indicative of crystalline twist angle, and using the rocking curve value from the symmetric case the twist angle was calculated to be 601 arcsec (0.1669°) [39]. In order to observe any possible rotational domains (twin planes) a pole figure was performed around the LiNbO$_3$ (012) reflection. Fig 4a shows threefold rotational symmetry and the absence of 60° rotational domains, indicating that any twin planes are minimized below the detection limit of the diffractometer, which is estimated at approximately 1% rotational domains by volume. The epitaxial relationship of the LiNbO$_3$ film (subscript f) on sapphire substrates (subscript s) is $<0001>_f \parallel <0001>_s$ and $<10\bar{1}0>_f \parallel <10\bar{1}0>_s$, which was determined by XRD phi scans shown in Fig 4c. While both crystals belong to 3 m symmetry groups, rotated grains may also nucleate with the relationship $<0001>_f \parallel <0001>_s$ and $<10\bar{1}0>_f \parallel <11\bar{2}0>_s$.

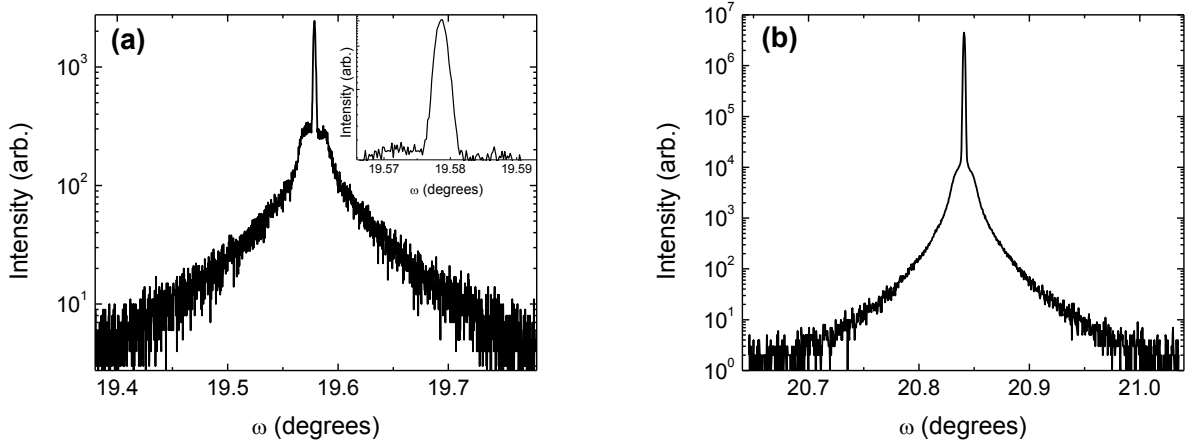

**Fig 3.** (a) Symmetric rocking curve of LiNbO$_3$ (006) on sapphire measured in triple-axis mode with a FWHM of 8.6 arcsec (0.0024°). The low intensity secondary peak was fit to a FWHM of 121 arcsec. **Inset** Enlarged view of the symmetric rocking curve maxima with ~30 data points indicating the measured value is well resolved. (b) Symmetric rocking curve of the substrate Al$_2$O$_3$ (006) peak showing coincident features for the film in (a). The FWHM of the narrow higher intensity section is 5.8 arcsec, while the lower intensity portion was fit to a FWHM of 67 arcsec.

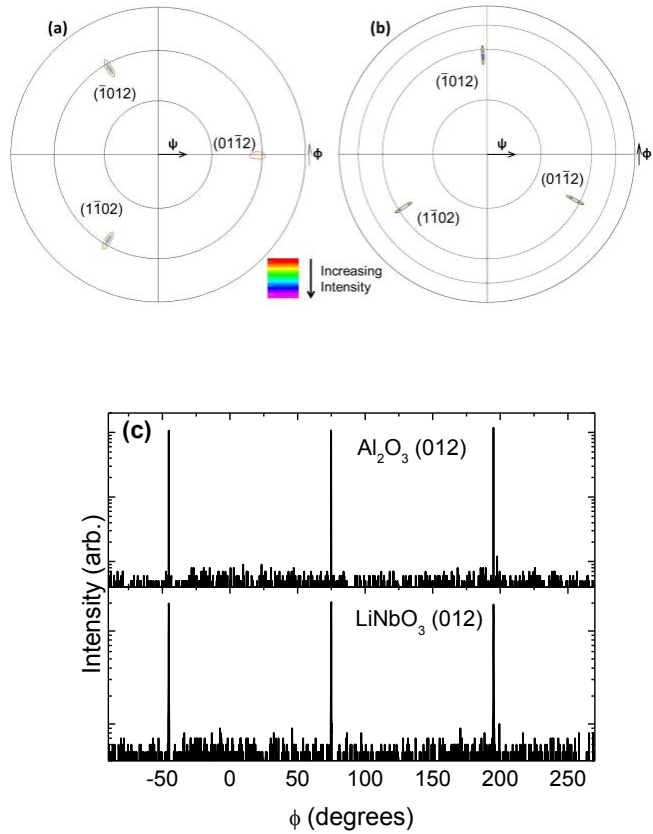

**Fig 4.** Pole figures around the LiNbO$_3$ (012) reflection on **(a)** sapphire and **(b)** LiTaO$_3$ substrates showing threefold symmetry. The absence of 60° rotational domains indicates the lack of twin planes. **(c)** XRD phi scans of the Al$_2$O$_3$ and LiNbO$_3$ (01$\bar{1}$2) reflections, respectively, showing the in-plane epitaxial relationship.

Twin plane suppression of lithium niobate on sapphire has been previously theorized from the standpoint of energetics [21,22]. A slightly energetically unfavorable cation alignment between Al, Li, and Nb exists at the interface, but it was shown that below 700℃ there is not enough thermally driven surface diffusion for all nucleation sites to form in the lowest energy configuration resulting in the growth of rotational domains. In MOCVD growth, which has produced some of the higher quality films, this lower limit for substrate temperature is a significant issue due to undesirable gas phase reactions which occur above 650 – 700℃ and can degrade the quality of the films as measured by symmetric XRD rocking curve [21]. By growing these MBE films at 1000℃, twin planes are eliminated.

The surface morphology of thin lithium niobate on sapphire was studied by AFM and is shown Fig 5. Non-coalesced columnar grains are sparsely nucleated with voids extending to the substrate. The plateau-like tops of the columnar grains are atomically flat with a measured surface roughness 0.1–0.2 nm RMS. Sparse grains like those observed in the present films have also been observed in materials grown by PLD and CVD [17,21]. Grain growth and coalescence was, therefore, studied as a function of film thickness and is depicted in Fig 5.

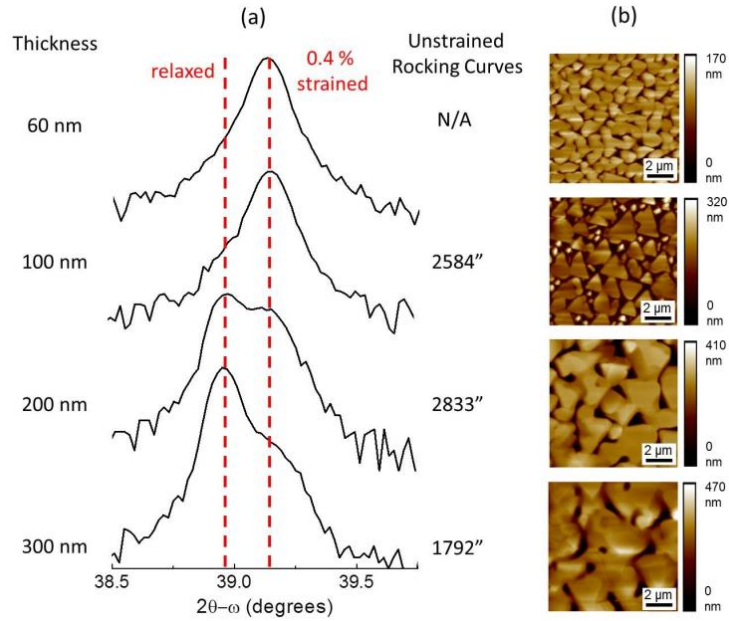

**Fig 5.** The evolution of relaxation and grain size/coalescence as a function of thickness is investigated by **(a)** XRD and **(b)** AFM. Relaxation, as shown by XRD 2θ-ω scans in (a), increases as the film thickness increases. Grain size also increases as a function of thickness, shown by AFM in (b), with incomplete coalescence at 300 nm. The unstrained rocking curve values (2θ = 38.98°), shown on the right of (a), indicate the onset of relaxation between 60 and 100 nm with the creation of defects and the subsequent annihilation of those defects as thickness increases.

Two phenomena are observed as the film thickness increases. Morphologically, the grain size increases and cracks appear as the films thicken, observed by AFM and shown in Fig 6. Structurally, film relaxation through defect formation is observed. The diffractograms in Fig 5a show a shift from 2θ = 39.15°, corresponding to 0.43% tensile strain as expected from the CTE mismatch, to relaxed lithium niobate at 2θ = 38.98° as the film thickness increases. Fig 5b shows the observed grain size and coalescence as measured by AFM. It was also observed that the symmetric rocking curve FWHM of the unstrained (006) reflection was initially very wide and subsequently narrows as the film thickness increases. This trend, shown in the right of Fig 5a, likely corresponds to the creation of defects upon film relaxation and subsequent partial annihilation of those defects. As shown in Fig 6a, it was observed that increasing the substrate temperature from 1000°C to 1025°C lead to a ~60% decrease in nuclei density and ~35% increase in grain size. These two general features, the inverse relationship between nuclei density and substrate temperature and the direct relationship between grain size and temperature, are common in epitaxy and has been observed before in the CVD growth of lithium niobate on sapphire [21,22].

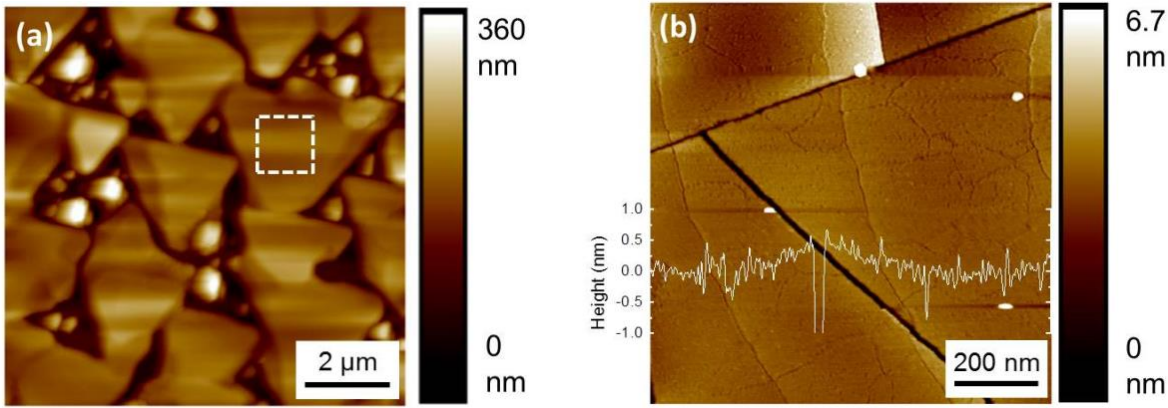

**Fig 6. (a)** AFM determined surface morphology of a 160 nm thick LiNbO$_3$ film grown on sapphire at 1025°C showing increased grain size and decreased nuclei density. The dotted box shows the scan location of **(b)** 1 μm square AFM image of the same film in (a) showing an atomically smooth surface with cracks. A line scan is overlaid to show the surface height in nanometers, confirming the atomically smooth surface of the plateau. The plateau surface roughness is 0.676 nm RMS.

*B. Lattice Matched Substrate*

Lithium niobate films were also grown on c-plane lithium tantalate which is lattice matched (0.12% mismatch) and isostructural. Films were grown under similar conditions to the films grown on sapphire at 1000°C. In order to avoid cracking due to thermal expansion, samples must be mounted in housings with significant expansion room (≫200 μm for a 1 cm square sample) and loose mechanical retaining clips to decrease the force on the sample surface [40] and transitioned at slow temperature ramp rates ≤20°C/min. Fig 7 shows the x-ray diffractogram of the lithium niobate on lithium tantalate with a peak value of 2θ = 38.999°, nearly identical to bulk LiNbO$_3$ (2θ = 38.98°). The diffractogram shows Pendellösung fringes which indicate a thickness of 112 nm and a smooth interface and surface. The inset of Fig 7 shows the symmetric rocking curve of the (006) lithium niobate reflection which has a FWHM of 193 arcsec. This value is nearly identical to the rocking curve value measured for the lithium tantalate substrate, 194 arcsec. Additionally, a pole figure shown in Fig 4b around the (012) lithium niobate reflection shows no twin planes as is the case for the films grown on sapphire.

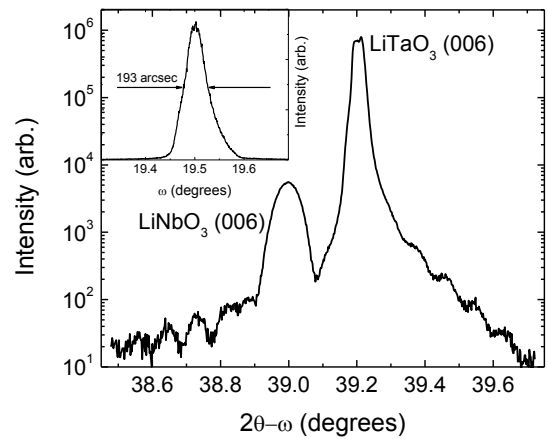

**Fig 7.** X-ray diffractogram of c-oriented LiNbO$_3$ grown on c-oriented LiTaO$_3$ with a peak location identical to the bulk value. Pendellösung fringes indicate a thickness of 112 nm. **Inset** Symmetric rocking curve at the LiNbO$_3$ (006) reflection with a FWHM of 193 arcsec, similar to the measured rocking curve of the underlying substrate which was 194 arcsec.

In contrast to the lithium niobate on sapphire samples, lithium niobate grown on lithium tantalate has a fully coalesced atomically smooth surface, shown in Fig 8 with a surface roughness of 0.175 nm RMS.

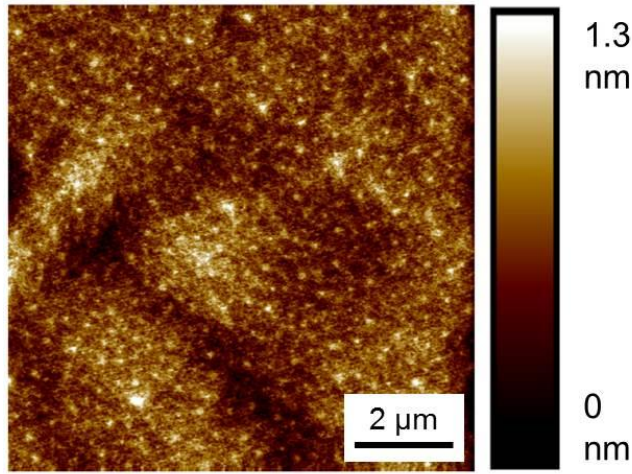

**Fig 8.** Surface morphology of 112 nm thick lithium niobate grown on lithium tantalate with a roughness of 0.175 nm RMS.

IV. CONCLUSION

High crystalline quality lithium niobate has been grown by MBE on lattice mismatched c-plane sapphire and lattice matched c-plane lithium tantalate. Films grown on c-plane sapphire show narrow symmetric rocking curves at the limit of XRD detection with a FWHM of 8.6 arcsec, equaling the tilt uniformity of bulk material and approaching that of the substrate measured as 5.6 arcsec. A pole figure around the $LiNbO_3$ (012) reflection in a skew geometry shows that there are no detectable 60° rotational domains in the material. The rocking curve shows a wider feature at lower intensity which corresponds to a similar feature in the substrate rocking curve showing the film is matching the slight tilt non-uniformity in the substrate. The rocking curve lower intensity portion for the film was fit to a FWHM of 121 arcsec, while that of the substrate was fit to a FWHM of 67 arcsec. AFM shows the $LiNbO_3$ grains are not fully coalesced even at 300 nm thickness. The morphology is highly columnar with voids extending to the substrate and plateaus that are atomically flat. The growth of lithium niobate on sapphire is shown to follow standard epitaxial trends regarding nucleation where increasing the substrate temperature leads to lower nuclei density and larger individual grains for a given thickness.

Lithium niobate grown on lattice matched lithium tantalate exhibits high structural and morphological quality on the same order as the substrate used. XRD analysis shows that the lithium niobate (006) symmetric peak is 193 arcsec wide at half maximum and the substrate peak is approximately the same at 194 arcsec wide. Films on lithium tantalate are also free of rotational domains, and the film surface is atomically flat on a 10 μm scale with a surface roughness of 0.175 nm RMS. This rotational disorder free, low XRD figure of merit material may be suitable for future thin film optics research.

ACKNOWLEDGEMENT

This work was supported by the Defense Threat Reduction Agency (DTRA), Basic Research Award # HDTRA-1-12-1-0031 administered by Dr. Jacob Calkins.

REFERENCES

1. R. Weis and T. Gaylord, Appl. Phys. A Mater. Sci. Process. **37**, 191 (1985).

2. K. K. Wong, *Properties of Lithium Niobate*, EMIS datareview series no. 28 (Institution of Electrical Engineers, and INSPEC, London, 2002).
3. D. R. Morgan, Ultrasonics **11**, 121 (1973).
4. R. V. Schmidt, I. P. Kaminow, and J. R. Carruthers, Appl. Phys. Lett. **23**, 417 (1973).
5. E. L. Wooten, K. M. Kissa, a. Yi-Yan, E. J. Murphy, D. a. Lafaw, P. F. Hallemeier, D. Maack, D. V. Attanasio, D. J. Fritz, G. J. McBrien, and D. E. Bossi, IEEE J. Sel. Top. Quantum Electron. **6**, 69 (2000).
6. M. N. Armenise, IEE Proc. J Optoelectron. **135**, 85 (1988).
7. M. Houe; P. D. Townsend, Appl. Phys. 28 **1747**, 1747 (1995).
8. G. Poberaj, H. Hu, W. Sohler, and P. Günter, Laser Photonics Rev. **6**, 488 (2012).
9. T. Kanata, Y. Kobayashi, and K. Kubota, J. Appl. Phys. **62**, 2989 (1987).
10. N. Fujimura, T. Ito, and M. Kakinoki, J. Cryst. Growth **115**, 821 (1991).
11. M. Shimizu, Y. Furushima, T. Nishida, and T. Shiosaki, Jpn. J. Appl. Phys. **32**, 4111 (1993).
12. P. R. Meek, L. Holland, and P. D. Townsend, Thin Solid Films **141**, 251 (1986).
13. J. J. Kingston, D. K. Fork, F. Leplingard, and F. A. Ponce, Mater. Res. Soc. Symp. Proc. **341**, 289 (1994).
14. S. Ono, O. Bose, W. Unger, Y. Takeichi, and S. Hirano, J. Am. Ceram. Soc. **81**, 1749 (1998).
15. Y. Shibata, K. Kaya, K. Akashi, M. Kanai, T. Kawai, and S. Kawai, J. Appl. Phys. **77**, 1498 (1995).
16. S.-H. Lee, T. K. Song, T. W. Noh, and J.-H. Lee, Appl. Phys. Lett. **67**, 43 (1995).
17. F. Veignant, M. Gandais, P. Aubert, and G. Garry, J. Cryst. Growth **196**, 141 (1999).
18. K. Matsubara, S. Niki, M. Watanabe, P. Fons, K. Iwata, and a. Yamada, Appl. Phys. A Mater. Sci. Process. **69**, S679 (1999).
19. H. Tamada, A. Yamada, and M. Saitoh, J. Appl. Phys. **70**, 2536 (1991).
20. S. Miyazawa, S. Fushimi, and S. Kondo, Appl. Phys. Lett. **26**, 8 (1975).
21. S. Y. Lee and R. S. Feigelson, J. Cryst. Growth **186**, 594 (1998).
22. R. S. Feigelson, J. Cryst. Growth **166**, 1 (1996).
23. A. A. Wernberg, H. J. Gysling, A. J. Filo, and T. N. Blanton, Appl. Phys. Lett. **62**, 946 (1993).
24. A. Dabirian, S. Harada, Y. Kuzminykh, S. C. Sandu, E. Wagner, G. Benvenuti, P. Brodard, S. Rushworth, P. Muralt, and P. Hoffmann, J. Electrochem. Soc. **158**, D72 (2011).
25. R. A. Betts and C. W. Pitt, Electron. Lett. **21**, 960 (1985).
26. W. A. Doolittle, A. G. Carver, and W. Henderson, J. Vac. Sci. Technol. B Microelectron. Nanom. Struct. **23**, 1272 (2005).
27. W. Doolittle, A. Carver, W. Henderson, and W. Calley, ECS Trans. **2**, 103 (2006).
28. J. D. Greenlee, W. L. Calley, W. Henderson, and W. A. Doolittle, Phys. Status Solidi **9**, 155 (2012).
29. Z. Sitar, F. Gitmans, W. Liu, and P. Gunter, Mater. Res. Soc. Symp. Proc. **401**, 255 (1996).
30. X. Lansiaux, E. Dogheche, D. Remiens, M. Guilloux-viry, a. Perrin, and P. Ruterana, J. Appl. Phys. **90**, 5274 (2001).
31. W. E. Henderson, W. Laws Calley, A. G. Carver, H. Chen, and W. Alan Doolittle, J. Cryst. Growth **324**, 134 (2011).
32. M. B. Tellekamp, J. D. Greenlee, J. C. Shank, and W. A. Doolittle, J. Cryst. Growth **425**, 225 (2015).
33. K. K. Lee, G. Namkoong, S. M. Madison, S. E. Ralph, W. A. Doolittle, M. Losurdo, G. Bruno, and H. K. Cho, Mater. Sci. Eng. B Solid-State Mater. Adv. Technol. **140**, 203 (2007).
34. K. K. Lee, G. Namkoong, W. A. Doolittle, M. Losurdo, G. Bruno, and D. H. Jundt, J. Vac. Sci. Technol. B **24**, 2093 (2006).
35. Y. S. Kim and R. T. Smith, J. Appl. Phys. **40**, 4637 (1969).
36. E. R. Dobrovinskaya, L. A. Lytvynov, and V. Pishchik, *Sapphire: Material, Manufacturing, Applications* (Springer US, 2009).
37. Z. Li and R. C. Bradt, J. Appl. Phys. **69**, 863 (1986).
38. F. Campbell, ASM Int. **#05224G**, 672 (2008).
39. S. R. Lee, A. M. West, A. A. Allerman, K. E. Waldrip, D. M. Follstaedt, P. P. Provencio, D. D. Koleske, and C. R. Abernathy, Appl. Phys. Lett. **86**, 1 (2005).
40. K. Lee, Implementation of AlGaN/GaN Based High Electron Mobility Transistor on Ferroelectric Materials for Multifunctional Optoelectronic-Acoustic-Electronic Applications, Georgia Institute of Technology, 2009.